\documentstyle[11pt,aaspp4]{article}

\def\etal{{\it et~al.}}
\def\ie{i.e.,}
\def\pc{C\^ot\'e}

\slugcomment{to appear in AJ}

\lefthead{}
\righthead{Globular Cluster Systems {\tt I.}}

\begin{document}

\title{Globular Cluster Systems {\rm I}: $V-I$ Color Distributions}

\author{Karl Gebhardt \altaffilmark{1,3}}
\affil{University of California Observatories / Lick Observatory, 
University of California, Santa Cruz, CA 95064}
\affil{Electronic mail: gebhardt@ucolick.org}

\author{Markus Kissler-Patig \altaffilmark{2}}
\affil{European Southern Observatory, Karl--Schwarzschild--Str.~2, 85748
Garching, Germany}
\affil{University of California Observatories / Lick Observatory, 
University of California, Santa Cruz, CA 95064}
\affil{Electronic mail: mkissler@eso.org}

\altaffiltext{1}{Hubble Fellow} 
\altaffiltext{2}{Feodor Lynen Fellow of the Alexander von Humboldt Foundation} 
\altaffiltext{3}{Guest User, Canadian Astronomy Data Centre, which is
operated by the Herzberg Institute of Astrophysics, National Research
Council of Canada}

\begin{abstract}

We have compiled data for the globular cluster systems of 50 galaxies
from the HST WFPC2 archive, of which 43 are type S0 or earlier. In
this paper, we present the data set and derive the $V-I$ color
distributions. We derive the first four moments of the color
distributions, as well as a measure for their non--unimodality.  The
number of globular clusters in each galaxy ranges from 18 (in
NGC~2778) to 781 (NGC~5846). For those systems having more than 100
clusters, seven of sixteen (44\%) show significant
bimodality. Overall, roughly half of all the systems in our sample
show hints of a bimodal color distribution. In general, the
distributions of the faint galaxies are consistent with unimodality,
whereas those of the brighter galaxies are not. We also find a number
of systems with narrow color distributions---with both mean red and
blue colors---suggesting that systems exist with only metal--rich or
only metal--poor globular clusters. We discuss their possible origins.

In comparing the moments of the $V-I$ distributions with various
galaxy properties for the early-type galaxies, we find the following
difference in the correlations between the field and cluster galaxy
populations: the peak $V-I$ color of the globular cluster distribution
correlates well with the central velocity dispersion---and hence the
Mg2 index and total luminosity---for galaxies in cluster environments;
there exists no such correlation for field galaxies. This difference
between cluster and field galaxies possibly reflects different
formation scenarios for their globular cluster systems. Among the
explanations for such a correlation, we consider either a larger age
spread in the field populations or the possibility that cluster
galaxies are always affected by significant accretion whereas some
field galaxies could host pure ``in situ'' formed populations.

\end{abstract}


\section{Introduction}

In the past decade, globular cluster systems were established as
powerful probes for the formation and evolution of galaxies
(e.g.~Harris 1991 and Ashman \& Zepf 1998 for reviews). With very few
exceptions, globular clusters exist in large numbers around all
galaxies. They are not only easy to detect; their properties are
readily influenced by the formation and evolution of their host
galaxies and, in a larger sense, their environment. Globular cluster
formation is also closely related to the star--formation history of
the galaxies (e.g.~Kissler-Patig~\etal\ 1998a). The study of globular
cluster systems, combined with the predictions of their properties
from various galaxy formation scenarios, can therefore be used to
constrain galaxy formation and evolution.

In recent years, tens of galaxies have been studied for their globular
cluster population (e.g.~compilation in Ashman \& Zepf 1998).
However, no systematic study of the properties of globular cluster
systems, including more than a dozen hosts, has been carried out.
Motivated by the need for a homogeneous study of a large number of
systems, we have exploited the Archive of the Hubble Space Telescope
as an ideal starting point for such a study, since it includes a large
number of images centered on galaxies in different filters and taken
in nearly identical conditions, leading to homogeneous photometry.

We begin by studying the properties of globular cluster systems around
a large number of {\it early--type} galaxies, then describe the
resulting sample in Sect.~2 and the pipeline reduction in Sect.~3.  We
study the color distributions of the systems (presented in Sect.~4 and
5), followed by a discussion of the properties of the color
distributions, correlations among these properties, and their relation
to the properties of the host galaxies (Sect.~6 and 7). Because color
denotes a mix of age and metallicity of the systems, globular cluster
systems have gained considerable interest after the discovery of
multi--modal distributions in some systems (Zepf \& Ashman 1993). Such
a system implies that at least in some galaxies, globular clusters
formed from several different mechanisms and possibly at several
epochs. Color distributions alone are thus already a strong diagnostic
for the history of the globular clusters systems and, by implication,
for the history of the host galaxies.  We discuss the properties of
the color distributions of our sample for various formation scenarios
in Sect.~8. In a companion paper, we use these data to constrain the
models of \pc~\etal\ (1998).

\section{The Galaxy Sample}

We searched the HST archive for all early-type galaxies that have been
observed in two colors with adequate exposure times, using only WFPC2
data from a variety of programs, with the Nuker Group (PI: Faber and
Richstone) providing most of it. 90\% of the galaxies where observed
in F555W ($V$) and F814W ($I$); some galaxies were observed through
the F547M and the F450W filters which we transformed to $V$ (see
below).  We did not use data taken in other filters. Although we may
have some some early-type galaxies, we attempted to make a complete
study from the archive. Furthermore, some of the galaxies below are
not early-types; we include them only for comparison, not for use in
the analysis.

Table 1 presents the data properties for each galaxy: galaxy name
(Col.~1), filter (Col.~2), total exposure time (Col.~3), number of
individual exposures (Col.~4), and PI name and number (Col.~5). Three
galaxies have more than one pointings and we include their total
exposure times from all pointings. There are 50 galaxies in
total. Table~2 summarizes the physical properties for the 43 galaxies
which are S0 or earlier.


\section{The Pipelined Reduction}

We reduced all of the data starting from the individual exposure. The
CADC provided the recalibrated science exposure (see
cadcwww.dao.nrc.ca for details of the recalibration). All data were
sent through a pipelined reduction. We divide the reduction into two
steps: the combination of the individual images, and the photometry
for the globular clusters. The first step requires shifting the
images, scaling by exposure times, and combining. We use the HST
header parameters for the telescope pointing (CRVAL1 and CRVAL2) to
determine the shift. Instead of using the telescope pointing, we could
have used reference objects in the images to measure the shift; for a
handful of galaxies, we confirmed that the telescope pointing provides
the same shift as when we used reference objects. We use a linear
shift so as not to smear cosmic-ray affected pixels, then scale the
images to a common exposure time. Different gains were used between
galaxies and even among different exposures of the same
galaxy. Finally, we transformed all images to gain 7. The combining
process must be robust, as many of the galaxies have a limited number
of individual exposures that make it difficult to remove cosmic-rays
adequately. We therefore use a biweight estimate for the
combining. The biweight attempts to estimate the mode of a
distribution by using an iterative approach whereby deviant points are
subsequently de-weighted. The parameters for the weighting result from
numerical simulations (see Beers~\etal\ 1991 and reference therein for
a fuller explanation). Unfortunately, some galaxies only have two
exposures in a given filter. To help decide whether a given pixel is
an outlier, we include both filters in the combining process. We
exclude the flagged pixels when combining in a given filter.

The second part of the reduction includes the photometry. Most of
these data have the center of the galaxy in one of the WFPC2 chips
(most are in the PC). To measure with greater reliability those
clusters which have the high-background galaxy under them, we first
subtract the galaxy light using a ``median-filtering'' technique;
instead of the median, we use a biweight estimate (explained above)
for the local flux value. We choose the parameters for the
filtering---\ie\ the inner radius equals 8 pixels and the outer radius
equals 14 pixels---based on results from experiments with no
background present.  In addition, we make an initial pass to detect
the globular clusters, then flag those pixels under the clusters to
make sure the filtering technique does not include them when
determining the galaxy component. We subtract the galaxy only in the
chip that contains the galaxy center.

We use SExtractor (Bertin \& Arnouts 1996) for the photometry, setting
the finding parameters to 2 connected pixels 2.5 $\sigma$ over the
local background as defined in a 32$\times$32 pixels box.  The frames
were further convolved with the PSF kernel given in the WFPC2 handbook
for the respective chips in order to optimize the finding (see
SExtractor manual, Bertin \& Arnouts 1996).

We must correct the resulting magnitudes since the photometry is
measured in a 2 pixel radius aperture. We correct the resulting
magnitudes to a 0.5\arcsec\ radius aperture, using the values proposed
in Whitmore~\etal\ (1997), and by 0.1 magnitude in order to
extrapolate to the total light (see Holtzman~\etal\ 1995). Finally,
the magnitudes measured in the HST filter set were converted into the
Johnson--Cousins $V$ and $I$ system by applying the transformations
given in Holtzman~\etal\ (1995). For the F555W and F814W filters used
in the vast majority of our data sets, empirical transformations
exists. For the F547M filter, we used the theoretical transformation
into the $V$ filter. Finally, we used the empirical transformation to
convert the F450W filter, used for the Fornax galaxies, into a Johnson
$B$ filter, followed by a linear transformation from $B-I$ into $V-I$
derived from the populations synthesis models of Bruzual \& Charlot
(1993) for ages $>$ 5 Gyrs and the full range of available
metallicities. The linear transformation follows $V-I=0.466~(B-I) +
0.14$. The average uncertainty on the magnitudes is 0.06; this
uncertainty scales with magnitude roughly so that at $V$=24 we have an
uncertainty of 0.1, and at $V$=20 an uncertainty of 0.01. We go down
to around 25.5 in $V$ and 24.5 in $I$, but we are only able to
approximate both our limiting magnitude and uncertainty, given that
the galaxies have a range of exposure times.

We note that the extrapolations from 2 pixel apertures to total
magnitudes, derived for point sources, are extremely uncertain when
applied to slightly extended objects such as globular clusters and can
significantly affect the final magnitudes (see Kundu \& Whitmore 1998,
Puzia~\etal\ 1999). However, the error of the extrapolation is similar
in the F555W and F814W filters, so that the final colors are not
affected by systematic errors of less than 0.03 magnitudes.

Finally, the data were corrected for foreground reddening using the
maps of Schlegel~\etal\ (1998), and the law E$(V-I)=1.55\cdot(B-V)$.


\section{The $V-I$ color distribution of the systems}

SExtractor provides an estimate of the magnitude, magnitude error,
ellipticity, FWHM, and star/galaxy classification. From these
measurements we select globular clusters from the objects that have
the following criteria: a magnitude error smaller than 0.1; $V-I$
color between 0 and 2; ellipticity smaller than 0.5; FWHM between 1
and 4 pixels (on both the PC and WF chips); and `non-star'
classification (set to reject remaining cosmic rays and hot pixels,
rather than to try to exclude point sources). No cut on the magnitude
exists since it correlates with the magnitude error. The number of
true globular clusters excluded by these criteria is typically less
than 10\%, somewhat higher for the most distant galaxies and those
with short exposure times. The contamination by foreground stars and
background galaxies passing our selection criteria is roughly constant
and estimated from the Hubble Deep Field North (Williams \etal
1996). The percentage of contamination depends on the distance and the
richness of the globular cluster system considered, and lies around
5\% for a typical system with $>100$ clusters at a distance of 10 to
20 Mpc. It can rise up to 30\% to 50\% for systems at significantly
larger distances---that we then excluded in the following analysis
where mentioned---or for systems with significantly less clusters.

Fig.~1 plots $V-I$ histograms for all of the 50 galaxies in the
sample. As a histogram is the crudest form of density estimation, we
also include the adaptive kernel density estimate as the solid line. A
histogram is a density estimate that utilizes both discreet and
equal-sized windows; it is therefore dependent on both the choice of
the window size and the location of the window edges. More robust
techniques include both weights for an individual datapoint's
contribution to a density estimate, and varying size that preserves
the noise properties (\ie\ sparse regions require larger binning). The
adaptive kernel estimate contains these properties, provides a
non-parametric density estimate on a grid. We refer to Silverman
(1986) for a complete discussion of the adaptive kernel estimate. It
works as follows: we obtain an initial density estimate, then adjust
the window width at each grid element according to the initial density
in such a way that low density regions have a large window width. We
use the Epanechnikov kernel, an inverted parabola, for the density
estimate.  The result of an adaptive kernel estimate can be highly
dependent on the initial smoothing, especially for small samples like
ours with $N<100$; thus, we use least-squares cross-validation to
choose the initial smoothing (see Silverman 1986, p. 48). This
technique minimizes an estimate of the integrated square error given
by $\int (\tilde{f} - f)^2$, where $\tilde{f}$ is the density
estimated from the data---dependent on the window width (the smoothing
parameter)---and $f$ is the underlying function we are trying to
estimate. One point is removed from the sample, and we estimate the
density for a given smoothing from the $N-1$ data points. Repeating
this $N$ times and using the average provides an estimate of the
underlying function $f$. The optimal smoothing is then the one that
minimizes the estimated integrated square error.

The dotted lines in Fig.~1 represent the 90\% confidence bands
obtained through a bootstrap estimate, a technique that draws with
replacement from the original sample of $N$ points. One bootstrap
realization contains $N$ points, which are not necessarily unique. For
each realization we estimate the distribution function and repeat this
procedure 1000 times. At each grid element we then have a distribution
of points, from which we choose the 5\% and the 95\% values,
representing the 90\% confidence band. These confidence bands reflect
only the variance, not the bias, of the estimate. The bootstrap
procedure includes the measurement uncertainties implicitly because
they are reflected in the original estimate of the colors.

\subsection{Measuring the Moments of the Distributions}

Moments represent a simple quantification of the distributions to with
which we correlate other galaxy properties (discussed below). However,
due to small number statistics and contamination, we desire a more
robust estimate for the moments other than the classical estimates. We
therefore use the biweight estimate of location (1st moment) and scale
(2nd moment), Finch's asymmetry index (3rd moment), and the Tail Index
(4th moment). Beers~\etal\ (1991) describe the biweight estimators in
detail. The asymmetry index measures skewness by comparing gaps
between the ordered data points on both the right and left side of the
distribution. The gaps are weighted according to their rank order,
thereby devaluing the contribution from outliers. The Tail Index
measures the spread of the dataset at the 90\% level relative to the
spread at 75\%, and is normalized such that a Gaussian has an index
equal one. Bird \& Beers (1993) provide an explanation of the
asymmetry index and tail index, and show that they are superior
estimates compared to the classical skewness and kurtosis. Table~3
presents these robust moments and their 68\% confidence band. A
bootstrap procedure, the same as described above, provides the
confidence bands.  The galaxies listed in italics in Table~3 are
late-type.


\section{Relations between Moments and Galaxy Properties}

In the following sections, we investigate the relations among the
properties of the globular cluster color distributions and their
relation to the properties of the host galaxies.  We restrict
ourselves to a homogeneous sample of globular cluster systems with
best determined properties.  We include S0's and ellipticals, but
exclude the late--type galaxies listed in italics in Table~3. We
further restrict the sample to galaxies with redshift less than 3000
km s$^{-1}$ (excluding NGC 1700 and NGC 4881), since at anything
smaller than this distance contaminating fore-/background objects
present no problem. We set a lower limit of the number of globular
clusters of 50, for which the moments are well defined. Finally, we
exclude the systems that were not observed in the F555W and F814W
filters to avoid systematic effects in the conversion to a homogeneous
photometric system (excluding NGC~4374, NGC~1399, NGC~1404).

The restricted sample includes 26 globular cluster systems for which
the distributions of properties are shown in Figure~2. Figure~2
presents all properties plotted against each other, as we discuss
below. Examining the range of the $V-I$ moments shows a relatively
sharp cut--off in the peak color at $V-I\simeq1.12$. The width or
scale appears to be around 0.16 for most systems, including an
artificial broadening by photometric errors of around 0.06 typically,
with only few systems having significantly broader distributions. Most
of the systems are slightly positively skewed---\ie\ they have the
tendency to cut--off more sharply in the blue and extend somewhat
towards the red. Finally, the tail index implies that most of the
distributions have tail weight consistent with a Gaussian
distribution.

\subsection{Peak to the 2nd, 3rd, and 4th Moments}

We show the relation between the peak and the scale (width) of the
distributions in the first plot on the upper line in Figure~2; an
enlarged version appears in Figure 3. We first notice that a blue
cut-off is seen in all distributions, implying a lower metallicity
limit for all globular clusters.  This limitation imposes a lower
cut--off in color and forbids a broadening of any distribution towards
the extreme blue (about $V-I<0.8$). Therefore, in order to have a blue
mean color, systems both must lack those red globular clusters that
could not be compensated by extreme blue clusters and must be
narrow. Interestingly, such systems with very little red clusters are
observed in our sample, i.e.~seem to exist among early--type galaxies.

On the other hand, red systems show a wide range of scale values,
indicating that both systems with {\it only} red clusters (red peak,
small scale) and systems with a wide range of globular cluster colors
exist. Figure~3 shows that the reddest systems (about $V-I>1.1$) have
systematically large scales. These systems must have a broad range of
colors, including blue clusters. For systems with $V-I>1.1$, a large
scale is partly expected due to the fact that the $V-I$ color changes
faster with metallicity at high metallicities
(e.g.~Kissler-Patig~\etal\ 1998b, compare with the theoretical
predictions, e.g.~Worthey 1994) and thus broadens any distribution.

The comparisons of the peak color to either the skewness or the tail
indicate no significant trends. The exception is expected: if a blue
system is strongly skewed, the skewness is towards the red color, and
vise-versa. However, not only do most systems fail to show a strong
skewness, the tail weights of the color distributions do not deviate
much from a Gaussian and show no systematic trends.

\subsection{Properties of the host galaxies}

For the host galaxies we compiled a number of properties to correlate
with the properties of the globular cluster color distribution. These
galaxy properties are given in Table 2, together with the source. The
galaxy properties include type, total luminosity, total color, central
velocity dispersion, central Mg2 index, environment density, isophotal
shape parameter, and, when applicable, effective radius.

The distribution of the properties for the 26 galaxies appear in the
plots in the lower-right side of Figure~2. The vast majority are
elliptical galaxies. We separate field from cluster galaxies at a
density parameter of $\rho =1$: \ie\ galaxies in groups such as Leo
(NGC~3377, NGC~3379, NGC~3384) will belong to the field, while
galaxies in small or medium sized clusters, such as Fornax and Virgo,
will belong to the cluster category.  We thus have 15 galaxies
belonging to the field (the green triangles in Figure~2) and 11
belonging to the cluster category (the blue circles).

While the Mg2 -- $\sigma$ relation is well defined for our galaxies,
the Faber--Jackson relation shows significant scatter, probably due to
noise in the distances.  Therefore, we use the central velocity
dispersion as a distance independent measure for the relative galaxy
mass, rather than the absolute magnitude. We have also made comparison
using log$(r_e~\sigma^2)$, which should reflect the galaxy mass better
than the dispersion alone, and find no better correlations.

\subsection{Correlations with the Whole Sample}

We first consider the whole sample, independent of environment.  No
significant correlation between the peak or scale of the globular
cluster systems and any galaxy property could be detected, as measured
by the Spearman rank-order correlation coefficients. We only note that
boxy galaxies ($a/a4 <0$) have systematically red peaks, and that the
peak values scatter most for intermediate galaxies when plotted
against $M_V$, $\sigma$, or $Mg2$ of the host galaxies.

At face value, this result might contradict the known dependence of
the mean metallicity of globular cluster systems from galaxy
luminosity (e.g.~Harris 1991, Forbes~\etal\ 1997), in so far as
brighter galaxies host globular clusters with a higher mean
metallicity. Therefore, we could have expected a systematically redder
peak with increasing $M_V$ or $\sigma$ of the host galaxy. But Ashman
\& Zepf 1998 noticed that the mean color of the clusters does not
correlate with the galaxy luminosity for early--type galaxies. In
fact, the relation between mean cluster metallicity and host galaxy
luminosity (cf.~Harris 1991) results from the fact that dwarf galaxies
possess less metal--rich globulars than spirals, which in turn host
less metal--rich globulars than giant early--type galaxies. Within a
given galaxy type, the relation between mean cluster metallicity and
galaxy luminosity is much less clear, if present at all.

\subsection{Comparing Field and Clusters Galaxies}

The recent models of hierarchical clustering formation for galaxies
(Kauffmann 1996, Baugh~\etal\ 1996) predict significant differences in
the formation and star--formation histories for field and cluster
galaxies. For this reason, we investigate these two classes separately
and split our samples of galaxies into field and cluster galaxies, as
described above.

According to a Spearman rank-order correlation coefficients, no
correlations are detected in the field sample of 15 galaxies. However,
for cluster galaxies, the peak values are found to correlate
positively with the host galaxy velocity dispersions at the $>99\%$
confidence level and, accordingly, at $>98\%$ confidence level with
their Mg2 index values.  Peak against galaxy velocity dispersion
appear in the seventh plot on the upper--line of Figure~2 for field
and cluster galaxies; an enlarged version is shown in Figure~3.  The
same results are true when using the total galaxy magnitude---\ie\
cluster galaxies show a correlation, whereas the field galaxies do
not.


\section{Bimodality}

Tests for bimodality are difficult to quantify. In many of the past
globular cluster systems studies, authors have used double Gaussian
fits to decide whether the system is bimodal. However, these results
are highly dependent both on the underlying assumption of Gaussianity
and on the criteria for deciding the number of sub-populations. More
general tests exist in the statistics literature that impose global
conditions on unimodality with no underlying assumptions about the
sub-populations. Here, we use the Dip statistic (Hartigan \& Hartigan
1985). The Dip test measures the maximum distance between the
empirical distribution function and the best fitting unimodal
distribution, with the significance level based on
simulations. Table~3 lists the value for the significance of the Dip
statistic, but not the Dip value itself. The same is true in Figure~2
and 4 where we plot the significance value only. A significance of 0.9
implies a 90\% probability that the distribution is not unimodal. We
do not calculate values below 0.5, as the Dip statistic is only
defined to discriminate against unimodality.

Figure~4 presents the DIP probabilities and confidence bands for all
of the galaxies. Out of 43 galaxies for which the dip can be measured,
9 galaxies show clear deviation from unimodality. These are NGC~1399,
NGC~1404, NGC~1426, NGC~3640, NGC~4472, NGC~4486, NGC~4526, NGC~4649,
and NGC~5846.  Six of these were already known from ground or space
studies to show several distinct populations (NGC~1399: Forbes~\etal\
1998, Ostrov~\etal\ 1998, Kissler-Patig~\etal\ 1997a; NGC~1404:
Forbes~\etal\ 1998; NGC~4472: Geisler~\etal\ 1996, Puzia~\etal\ 1999;
NGC~4486: Whitmore~\etal\ 1995, Elson \& Santiago 1996; NGC~4649:
Neilsen~\etal\ 1999; NGC~5846: Forbes~\etal\ 1997). Using a test to
discriminate between single and double Gaussians, Neilsen~\etal\ (1999)
found significant bi-modal population in four of their eight galaxies
studied. Using the same datasets, we confirm most of their
measurements of bi-modality; we disagree only in NGC~4552 where we
find no evidence for it.

A further eleven galaxies show DIP values greater than 0.5 at the 1 to
2 $\sigma$ level---\ie\ to have potential deviations from
unimodality. These are NGC~584, NGC~596, NGC~821, NGC~1023, NGC~2434,
NGC~2778, NGC~3115, NGC~3610, NGC~4365, NGC~4458, and NGC~4478. The
studies of Neilsen \& Tsvetanov (1999) and Kissler-Patig~\etal\ (1999)
show that NGC 4478 does not seem to have two clearly distinct
populations. In contrast, the recent study of Kundu \& Whitmore (1998)
reveals that NGC 3115 hosts two distinct populations. We speculate
that at least some of these systems might have several distinct
populations. Further detailed analysis are required for these
candidates.

The remaining 23 galaxies do not show deviations from unimodality. We
note that the detection of a dip does not correlate with the number of
globular clusters detected; the sample size is taken into account when
computing the significance of the statistic.

In Figure~2, the bimodality plots of interest are those which show DIP
significance versus both the absolute magnitude and the environment
density of the host galaxy. The detection of bimodality does not seem
to correlate with the environment. However, some trend can be seen
with absolute magnitude: low--luminosity systems ($M_V>-20$) appear
unimodal. This trend has already been noted by Kissler-Patig (1997) in
relation to other differences between low-- and high--luminosity
systems. It is not clear yet how much of these differences can be
associated with fundamental differences between the two types of
galaxies, and how much is just a bias in the observations.  Indeed, if
the blue peak is constant in all galaxies, while the red one scales
with galaxy luminosity, the two peaks will be harder to disentangle in
fainter galaxies (cf.~Kissler-Patig~\etal\ 1998a). On the other hand,
high--luminosity systems do not appear systematically to show
bimodality. We note the interesting cases of NGC~4406 (M86) and
NGC~4374, two giant ellipticals in Virgo. The first galaxy, a central
giant elliptical dominating an infalling group (Schindler \etal\ 1999), is
clearly unimodal. NGC~4374 also has a unimodal distribution; however,
it is heavily skewed to the red, so that a large number of red
(presumably metal--rich) globular clusters exist but seem to form a
continuous distribution rather than a distinct population.

In summary, it seems harder to rule out the presence of several
populations than to detect them. In our sample, roughly half of the
systems show multiple populations of globular clusters, independently
of the environment but preferentially in the brightest galaxies. Our
future challenge will be to confirm or rule out single populations in
low--luminosity systems.


\section{Combining Galaxy Samples}

The moments of the $V-I$ distributions are not well estimated in many
of our galaxies based on small numbers of globular clusters. To
compare correlations more robustly, we combine the different globular
cluster samples to give better estimates of the moments. Because we do
not have to exclude galaxies based on there being a small number of
measured globular clusters, we now have 43 galaxies in the combined
sample. We use the four main parent galaxy properties---magnitude,
$\sigma$, Mg2 index, and density environment---for this analysis,
breaking the sample into two bins. Figure~5 plots the results for the
combined $V-I$ distributions. Table~4 presents the various cuts used
for the combined samples and the moments and DIP probability for each
sub-sample. Table~4 also presents the number of galaxies and the
number of globular clusters in each sub-sample. Each sub--sample has
greater than 800 clusters where the noise from the moment estimation
is minimal. The choice for the cuts in the sub-samples results from
our inspection of the particular distribution; we attempted to find a
clear division between the samples, and, when none was apparent, we
used cuts which had roughly equal range among the particular
variable. The cuts are partly subjective, but the results do not
depend strongly on them.

The central location of the $V-I$ peak is statistically different only
in the division between faint and bright galaxies. The faint galaxies
have a peak around 0.94, and the bright galaxies have a peak around
1.13. For the other divisions, the $V-I$ peaks are similar, but the
tail indexes differ significantly. Either faint galaxies, low
$\sigma$, low Mg2, or low density cause the $V-I$ distribution to have
significantly more tail weight than compared to their counterpart
sample.

The most extreme differences between the combined samples results from
the DIP probability. Galaxies that are faint, have low $\sigma$ or low
Mg2 are consistent with a unimodal distribution, whereas their
counterpart samples are highly non-unimodal. This effect is also seen
when looking at individual galaxies (see Section~6 above), and it
appears to be a fundamental property of galaxies.


\section{Discussion}

\subsection{The Color Distributions}

An important discovery in the study of globular clusters systems in
the last decade is the fact that the globular clusters' color
distribution of many ellipticals show several peaks (Zepf \& Ashman
1993). This property has been interpreted as the presence of several
distinct populations of globular clusters around the host galaxy. The
origin of these different populations is still under debate and might
well have several causes. The original explanation argues that
spiral--spiral mergers provide old globular clusters from the
progenitors as well as produce a new population during the merger
(Ashman \& Zepf 1992, see also Schweizer 1987). However, the accretion
of enough dwarf galaxies with metal poor clusters will also obviously
build up a distinct metal--poor population. C\^ot\'e~\etal\ (1998),
and Hilker (1999) investigated this idea. Yet another process involves
the pre--galactic formation of metal--poor clusters as the building
blocks of galaxies, followed by the formation of more metal--rich
globular clusters during the collapse of the galaxy (see Kissler-Patig
1997, and Kissler-Patig~\etal\ 1998a). Only slightly different is the
theory that galaxies formed all their globular clusters {\it in situ}
and had two main phases of star formation (e.g.~Forbes~\etal\ 1997,
Harris~\etal\ 1998, Harris~\etal\ 1999). Today's challenge is to
discriminate between the various scenarios with detailed comparison to
the present data.

The diversity of the color distributions present in our sample
indicates that indeed the scenario is complex. We discuss in turn some
of the groups:

\vskip 5mm

$\bullet$ Galaxies that host ``only red clusters''---\ie\ those having
a red peak and a small scale value. The existence of a blue population
in these galaxies is not excluded but is unlikely to be
significant. The favored scenario for these galaxies is that a single
dominant collapse created a bulge and a red population of clusters.
We cannot rule the accretion scenario out if these galaxies were
formed in an environment poor in material suitable for accretion
(e.g.~dwarf galaxies, cf.~Chapelon 1999). These galaxies would then be
in low density environments or in the outskirts of clusters; neither
is confirmed nor ruled out by our present statistics. Finally, the
merger scenario could also work for these galaxies if the progenitors
were poor in blue clusters. If the merger happened at early times, the
progenitors could be still in a gaseous form, or, alternatively, could
be low--surface--brightness galaxies for late mergers in the field
that host a large amount of gas but apparently few clusters. However,
a scenario with gaseous mergers at early times is hard to distinguish
from the two-stage {\it in situ} scenario.

\vskip 5mm

$\bullet$ Galaxies that host ``only blue clusters''---\ie\ that have a
blue peak and a small scale value. These galaxies do not seem to host
a significant population of red (presumably metal--rich) clusters. In
this case, none of the above scenarios would easily be able to explain
such systems; one would have to advocate the formation of bulge stars
without the formation of clusters. More likely is a ``blue'' red peak,
either because the red clusters formed with a low metallicity or because
they have a significant younger age (about half as old as the true
blue clusters, cf.~Kissler-Patig~\etal\ 1998a) that make them appear
blue.  In these last two cases, all scenarios work, with some
preference to the merger scenario in the frame of hierarchical
clustering models that predict late mergers in the field.

\vskip 5mm

$\bullet$ Galaxies with broad or bimodal color distributions---\ie\
that which presumably host a comparable amount (say within a factor of
about 3) of blue and red clusters. These are the cases for which the
scenarios described above were designed, and we refer to the
respective papers for the pros and cons.

\vskip 5mm

Finally, we note that as long as some information is missing (merger
epoch, globular clusters formation mechanisms, faint end of the
luminosity function in different environments, etc...), and given the
diversity of scenarios, it seems that not always the same mechanism is
the dominant one.

\subsection{Field and Cluster Systems}

Although based on small number statistics, field and cluster
early--type galaxies seem to differ in the properties of their
globular cluster systems. Whereas cluster galaxies show a correlation
between peak value and the galaxy size (as measured by the velocity
dispersion), such a relation must be ruled out for field galaxies.

For cluster ellipticals, the recent spectroscopic studies
(Cohen~\etal\ 1998, Kissler-Patig~\etal\ 1998b) show that all globular
clusters are old (around 10 Gyr or older). At these ages, the clusters
can be considered coeval with respect to the influence of their ages
on color. Color differences can therefore be interpreted as almost
pure metallicity differences, and the relation between peak color and
galaxy luminosity might reflect the equivalent of the Mg2--$\sigma$
relation for galaxies.  Forbes~\etal\ (1997) suggest that in systems
with two distinct globular cluster populations, the mean metallicity
of the red clusters correlates with the luminosity of the host galaxy;
the blue peak does not. Their conclusion implies that the correlation
of the total sample's mean color with galaxy size is driven by the
metal--rich globular clusters only. Consequently, this would support
scenarios in which the metal--poor globular clusters formed in events
uncorrelated with their future host galaxy---e.g.,~in systems that are
separate entities from their future hosts. On the other hand, the same
research suggests that the metal--rich globular clusters formed in
events correlated with the future host galaxy---e.g.,~the formation of
a bulge in a collapse or merger event. In addition, this correlation
suggests that the blue sub--populations have roughly a constant color,
while for the red population both the peak shifts and the scale
increase as a function of color.

Perhaps surprisingly, the peak color of globular clusters in field
galaxies does not correlate with the velocity dispersion ($\approx$
size) of their host galaxy. This result appears to be in contrast with
the observation that field and cluster galaxies do exhibit the same
Mg2--$\sigma$ relations (Bernardi~\etal\ 1998). We see several
possible explanations.

A first alternative might be that the correlation for cluster galaxies
is artificial.  We verified that it is not caused by sampling
different regions in the different galaxies. Indeed, the ratio of blue
to red galaxies does vary with radius (Geisler~\etal\ 1997,
Kissler-Patig~\etal\ 1997b, Puzia~\etal\ 1999). Furthermore, the WFPC2
field samples the outer parts less in larger galaxies than in small
ones, so that the relative ratio of blue to red galaxies is
systematically smaller in large galaxies at fixed field of
view. Puzia~\etal\ (1999) tested the importance of this effect on NGC
4472, which has exposures centered on the galaxy as well as several
arcminute offsets north and south. The respective peaks of the blue
and red populations do not change with radius within the errors;
however, the varying ratio of blue to red causes the overall peak to
shift by 0.02 mag from the inner to the outer field. The effect is too
small, though, to explain any systematic correlation.  Moreover, in
our sample, field and cluster galaxies span different values of galaxy
$\sigma$. Where cluster galaxies are of intermediate to giant size,
the field galaxies span the low to intermediate size range.  If indeed
intermediate and giant early--type galaxies differ (see above,
cf. Kissler-Patig 1997), the presence/absence of a correlation would
be due not to environment, but rather to a function of the galaxy
size.  For example, one could speculate that small galaxies are not
surrounded by a ``halo'' of blue clusters, thus showing a small trend
of system peak with size. Intermediate size or larger galaxies might
or might not have such a halo of blue clusters, complicating the
interpretation of the peak color and introducing a scatter due to the
possible existence of blue clusters. Giant ellipticals would always
have blue clusters, and a significant redder population that
correlates with galaxy size, leading to an overall trend of peak color
with galaxy size again. In other words, the correlation would be an
artifact of the mix of populations which show a correlation with
environment, rather than a physical difference in the enrichment
histories.

A second alternative would be a physical interpretation of the result.
Thus, the presence of a Mg2--$\sigma$ relations in field and cluster
galaxies, but absence/presence respectively of a peak--$\sigma$
relation, would indicate that stars and globular clusters are coupled
in cluster galaxies but not in field galaxies. We regard this
alternative as unlikely, however, since cluster formation and star
formation appear to go together.

Finally, Mg2 and $V-I$ color could vary differently with age and
metallicity.  The variations of Mg2 with age and metallicity are still
uncertain and if, for example, the metal--rich population in field
ellipticals show a large spread in age from galaxy to galaxy, a trend
could mimic an Mg2--$\sigma$ relation in the field but not a
peak--$\sigma$ relation. In contrast, since all globular clusters in
cluster galaxies are old, such a trend would not be present, and both
Mg2 and peak would trace metallicity only. Clearly such speculations
require a much better understanding of the influence of age and
metallicity on the properties of composite stellar
populations. Detailed studies of the ages and metallicities of the
systems are needed.


\section{Summary and Outlook}

The results presented in this paper suggest that we can use globular
cluster systems to study the merger/formation history of both itself
and its host galaxy. The differences we see between the field and
cluster systems suggest that the cluster galaxies are more influenced
by accretion, whereas the field systems may be the result of ``in
situ'' globular cluster formation. Combining this result with the
strong signature of multi-modality seen in many of the systems
provides an observational comparison to the various formation
scenarios.  In a companion paper (\pc~\etal\ 2000), we will compare
these results to the theoretical predictions of \pc~\etal\
(1998). More detailed studies of the systems that are candidates for
bimodal distributions are underway, including follow--up
near--infrared photometry and multi--object spectroscopy.

All photometry results are available from the authors.

\acknowledgments

We are indebted to Pat \pc\ for his help and advises in some of
the early--stages of this study. Thanks to Thomas Puzia for his help
with the NGC 4472 data and extra--tests. We thank Alvio Renzini for
enlightening discussions, as well as the ENEAR team for providing
galaxy data prior to publication, especially Mariangela Bernardi for
her help in digging these latter out.  KG is supported by NASA through
Hubble Fellowship grant HF-01090.01-97A awarded by the Space Telescope
Science Institute, which is operated by the Association of the
Universities for Research in Astronomy, Inc., for NASA under contract
NAS 5-26555.  MKP was partly supported by a Feodor Lynen Fellowship of
the Alexander von Humboldt Foundation.

\clearpage

\begin{deluxetable}{lccrl}
\footnotesize
\tablewidth{0pt}
\tablecaption{Observational Parameters}
\tablehead{
\colhead{Galaxy}        &
\colhead{Filter}        &
\colhead{T$_{\rm exp}$} &
\colhead{N$_{\rm exp}$} &
\colhead{PI (number)}   }
\startdata
N0584    & F555W & 1520 &  5 & Faber (6099)     \nl
         & F814W & 1450 &  7 & Faber (6099)     \nl
N0596    & F555W & 1600 &  4 & Richstone (6587) \nl
         & F814W & 1800 &  3 & Richstone (6587) \nl
N0821    & F555W & 1520 &  5 & Faber (6099)     \nl
         & F814W & 1450 &  7 & Faber (6099)     \nl
N1023$^{\rm a}$& F555W & 2920 & 11 & Faber,Richstone (6099,6587) \nl
         & F814W & 3780 & 10 & Faber,Richstone (6099,6587) \nl
N1316    & F450W & 5000 &  5 & Grillmair (5990) \nl
         & F814W & 1860 &  3 & Grillmair (5990) \nl
N1399    & F450W & 5200 &  4 & Grillmair (5990) \nl
         & F814W & 1800 &  3 & Grillmair (5990) \nl
N1404    & F450W & 5000 &  5 & Grillmair (5990) \nl
         & F814W & 1860 &  3 & Grillmair (5990) \nl
N1426    & F555W & 1600 &  4 & Richstone (6587) \nl
         & F814W & 1800 &  3 & Richstone (6587) \nl
N1700    & F555W & 1600 &  4 & Richstone (6587) \nl
         & F814W & 1800 &  3 & Richstone (6587) \nl
N2300    & F555W & 1520 &  5 & Faber (6099)     \nl
         & F814W & 1450 &  7 & Faber (6099)     \nl
N2434    & F555W & 1300 &  2 & Carollo (5943)   \nl
         & F814W & 1000 &  2 & Carollo (5943)   \nl
N2778    & F555W & 1520 &  5 & Faber (6099)     \nl
         & F814W & 1450 &  7 & Faber (6099)     \nl
N3115    & F555W & 1015 &  6 & Faber (5512)     \nl
         & F814W & 1236 &  6 & Faber (5512)     \nl
N3377    & F555W & 1330 &  6 & Faber (5512)     \nl
         & F814W & 1236 &  6 & Faber (5512)     \nl
N3379    & F555W & 1660 &  4 & Faber (5512)     \nl
         & F814W & 1340 &  4 & Faber (5512)     \nl
N3384    & F555W & 1430 &  6 & Faber (5512)     \nl
         & F814W & 1290 &  6 & Faber (5512)     \nl
N3585    & F555W & 1200 &  6 & Richstone (6587) \nl
         & F814W & 1900 &  3 & Richstone (6587) \nl
N3610    & F555W & 1400 &  6 & Richstone (6587) \nl
         & F814W & 2100 &  3 & Richstone (6587) \nl
N3640    & F555W & 1200 &  6 & Richstone (6587) \nl
         & F814W & 2100 &  3 & Richstone (6587) \nl
N4125    & F555W & 1400 &  6 & Faber (6587)     \nl
         & F814W & 2100 &  3 & Faber (6587)     \nl
N4192    & F555W &  660 &  3 & Rubin (5375)     \nl
         & F814W &  660 &  3 & Rubin (5375)     \nl
N4291    & F555W & 1520 &  5 & Faber (6099)     \nl
         & F814W & 1450 &  7 & Faber (6099)     \nl
N4343    & F555W &  660 &  3 & Rubin (5375)     \nl
         & F814W &  660 &  3 & Rubin (5375)     \nl
N4365    & F555W & 2200 &  2 & Brodie (5920)    \nl
         & F814W & 2300 &  2 & Brodie (5920)    \nl
N4374    & F547M & 1200 &  2 & Bower (6094)     \nl
         & F814W &  520 &  2 & Bower (6094)     \nl
N4406    & F555W & 1500 &  3 & Faber (5512)     \nl
         & F814W & 1500 &  3 & Faber (5512)     \nl
N4450    & F555W &  520 &  3 & Rubin (5375)     \nl
         & F814W &  520 &  3 & Rubin (5375)     \nl
N4458    & F555W & 1340 &  4 & Faber (5512)     \nl
         & F814W & 1120 &  5 & Faber (5512)     \nl
N4472$^{\rm a}$& F555W & 4400 &  4 & Brodie (5920)    \nl
         & F814W & 4600 &  4 & Brodie (5920)    \nl
N4473    & F555W & 1800 &  3 & Faber (6099)     \nl
         & F814W & 2000 &  4 & Faber (6099)     \nl
N4478    & F555W & 1600 &  4 & Richstone (6587) \nl
         & F814W & 1800 &  3 & Richstone (6587) \nl
N4486    & F555W & 2810 &  6 & Macchetto (5477) \nl
         & F814W & 2430 &  5 & Macchetto (5477) \nl
N4486B   & F555W & 1800 &  3 & Faber (6099)     \nl
         & F814W & 2000 &  4 & Faber (6099)     \nl
N4526    & F555W &  520 &  3 & Rubin (5375)     \nl
         & F814W &  520 &  3 & Rubin (5375)     \nl
N4536    & F555W &  660 &  3 & Rubin (5375)     \nl
         & F814W &  660 &  3 & Rubin (5375)     \nl
N4550    & F555W & 1200 &  3 & Rubin (5375)     \nl
         & F814W & 1200 &  3 & Rubin (5375)     \nl
N4552    & F555W & 2400 &  4 & Faber (6099)     \nl
         & F814W & 1500 &  3 & Faber (6099)     \nl
N4569    & F555W &  520 &  3 & Rubin (5375)     \nl
         & F814W &  520 &  3 & Rubin (5375)     \nl
N4594    & F547M & 1340 &  4 & Faber (5512)     \nl
         & F814W & 1470 &  6 & Faber (5512)     \nl
N4621    & F555W & 1380 &  6 & Faber (5512)     \nl
         & F814W & 1280 &  6 & Faber (5512)     \nl
N4649    & F555W & 2100 &  2 & Westphal (6286)  \nl
         & F814W & 2500 &  2 & Westphal (6286)  \nl
N4660    & F555W & 1000 &  5 & Faber (5512)     \nl
         & F814W &  850 &  5 & Faber (5512)     \nl
N4881    & F555W & 7200 &  8 & Westphal (5233)  \nl
         & F814W & 7200 &  8 & Westphal (5233)  \nl
N5018    & F555W & 1200 &  6 & Richstone (6587) \nl
         & F814W & 1900 &  3 & Richstone (6587) \nl
N5061    & F555W & 1200 &  6 & Richstone (6587) \nl
         & F814W & 1900 &  3 & Richstone (6587) \nl
N5845    & F555W & 2140 &  5 & Faber (6099)     \nl
         & F814W & 1120 &  5 & Faber (6099)     \nl
N5846$^{\rm a}$& F555W & 6600 &  6 & Brodie (5920)    \nl
         & F814W & 6900 &  6 & Brodie (5920)    \nl
N7192    & F555W & 1300 &  2 & Carollo (5943)   \nl
         & F814W & 1000 &  2 & Carollo (5943)   \nl
N7457    & F555W & 1380 &  6 & Faber (5512)     \nl
         & F814W & 1280 &  6 & Faber (5512)     \nl
IC4889   & F555W &  600 &  1 & Carollo (5943)   \nl
         & F814W &  600 &  1 & Carollo (5943)   \nl
\enddata
\tablenotetext{a}{N1023 and N4472 have two different pointings; 
N5846 has three.}
\end{deluxetable}
 
\begin{deluxetable}{llccccrc}
\footnotesize
\tablecaption{Galaxy Properties}
\tablewidth{0pt}
\tablehead{
\colhead{Galaxy}       &
\colhead{Type}         & 
\colhead{$M_V$}        &
\colhead{log($\sigma$)}&
\colhead{Mg2}          &
\colhead{log($\rho$)}  &
\colhead{a4}           &
\colhead{log(R$_e$)(pc)}} 
\startdata
N584   & E4     & --21.16 &  2.34 &  0.29 &  0.42 &  1.50   &  3.51   \nl
N596   & Epec   & --20.93 &  2.22 &  0.25 &  0.40 &  1.30   & \nodata \nl
N821   & E6     & --20.84 &  2.32 &  0.32 &  0.08 &  2.50   &  3.45   \nl
N1023  & SB0    & --20.61 &  2.34 &  0.34 &  0.57 & \nodata &  3.26   \nl
N1399  & E1pec  & --21.97 &  2.52 &  0.37 &  1.59 &  0.00   & \nodata \nl
N1404  & E1     & --21.56 &  2.38 &  0.34 &  1.59 &  0.50   & \nodata \nl
N1426  & E4     & --20.42 &  2.18 &  0.28 &  0.66 & \nodata &  3.44   \nl
N1700  & E4     & --21.90 &  2.36 &  0.28 & \nodata &  0.90 & \nodata \nl
N2300  & SA0    & --20.76 &  2.42 &  0.33 &  0.14 & --0.60 &  3.58  \nl
N2434  & E0     & --20.60 &  2.31 &  0.27 &  0.19 & \nodata &  3.49  \nl
N2778  & E      & --19.43 &  2.26 &  0.34 &  0.28 & \nodata & \nodata  \nl
N3115  & S0     & --20.82 &  2.45 &  0.34 &  0.08 &  8.00 &  3.17  \nl
N3377  & E5     & --19.96 &  2.18 &  0.29 &  0.49 &  1.20 &  3.21  \nl
N3379  & E1     & --20.83 &  2.35 &  0.34 &  0.52 &  0.10 &  3.23  \nl
N3384  & SB0    & --19.89 &  2.21 &  0.29 &  0.54 & \nodata & \nodata  \nl
N3585  & E7     & --21.75 &  2.34 &  0.32 &  0.12 &  4.50 &  3.50  \nl
N3610  & E5     & --20.92 &  2.21 &  0.27 &  0.30 &  2.50 & \nodata  \nl
N3640  & E3     & --21.75 &  2.24 &  0.27 &  0.18 & --0.20 & \nodata  \nl
N4125  & E6pec  & --22.09 &  2.25 &  0.32 &  0.34 &  1.00 &  3.43  \nl
N4291  & E      & --20.69 &  2.43 &  0.32 &  0.36 & --0.40 &  3.64  \nl
N4365  & E3     & --22.12 &  2.42 &  0.35 &  2.93 & --1.10 &  3.79  \nl
N4374  & E1     & --22.21 &  2.44 &  0.32 &  3.99 & --0.40 & \nodata  \nl
N4406  & S0     & --22.41 &  2.40 &  0.33 &  1.41 & --0.70 & \nodata  \nl
N4458  & E0     & --19.06 &  2.02 &  0.25 &  3.21 & \nodata & \nodata  \nl
N4472  & E2     & --22.59 &  2.48 &  0.34 &  3.31 & --0.30 &  3.89  \nl
N4473  & E5     & --20.80 &  2.26 &  0.33 &  2.17 &  0.90 &  3.60  \nl
N4478  & E2     & --19.94 &  2.13 &  0.27 &  3.92 & --0.80 & \nodata  \nl
N4486  & E0     & --22.48 &  2.56 &  0.30 &  4.17 &  0.00 &  3.89  \nl
N4486b & cE0    & --17.75 &  2.30 &  0.30 &  4.17 & \nodata &  3.51  \nl
N4526  & SAB0   & --20.89 &  2.47 &  0.30 &  2.45 & \nodata &  3.57  \nl
N4550  & SB0    & --18.81 &  2.01 &  0.17 &  2.97 & --0.70 & \nodata  \nl
N4552  & E      & --21.26 &  2.42 &  0.35 &  2.97 & \nodata &  3.35  \nl
N4621  & E5     & --21.63 &  2.40 &  0.35 &  2.60 &  1.50 &  3.54  \nl
N4649  & E2     & --22.31 &  2.56 &  0.37 &  3.49 & --0.50 &  3.74  \nl
N4660  & E      & --19.26 &  2.29 &  0.33 &  3.37 &  2.70 &  3.56  \nl
N4881  & E      & --21.60 &  2.33 &  0.29 & \nodata & \nodata & \nodata  \nl
N5018  & E3     & --22.31 &  2.32 &  0.21 &  0.29 &  2.00 & \nodata  \nl
N5061  & E0     & --21.62 &  2.30 &  0.27 &  0.31 & --1.80 &  3.90  \nl
N5845  & E      & --19.63 &  2.42 &  0.32 &  0.84 &  0.80 & \nodata  \nl
N5846  & E0     & --22.23 &  2.40 &  0.34 &  0.84 &  0.00 &  3.37  \nl
N7192  & SA0    & --21.49 &  2.27 &  0.26 &  0.28 & \nodata &  3.71  \nl
N7457  & SA0    & --18.25 &  1.76 &  0.19 &  0.13 & \nodata & \nodata  \nl
IC4889 & E5     & --21.26 &  2.22 &  0.26 &  0.13 &  1.00 & \nodata  \nl
\enddata
\end{deluxetable}

\begin{deluxetable}{lcccccc}
\footnotesize
\tablecaption{Moments of $V-I$ Distributions}
\tablewidth{0pt}
\tablehead{
\colhead{Galaxy}      &
\colhead{N$_{\rm GC}$}& 
\colhead{$V-I$ peak}  &
\colhead{Scale}       &
\colhead{Skewness}    &
\colhead{Tail}        &
\colhead{DIP}         } 
\startdata
       N584 &   67 &  $1.092~_{1.062}^{1.123}$ & $0.159~_{0.130}^{0.192}$ & $-0.013~_{-1.159}^{+1.128}$ & $1.080~_{0.864}^{1.484}$ & $0.649~_{0.500}^{0.852}$ \nl
       N596 &   31 &  $1.049~_{1.004}^{1.096}$ & $0.158~_{0.115}^{0.201}$ & $+0.306~_{-0.981}^{+1.541}$ & $1.176~_{0.833}^{2.049}$ & $0.661~_{0.500}^{0.778}$ \nl
       N821 &   55 &  $1.008~_{0.976}^{1.041}$ & $0.142~_{0.126}^{0.158}$ & $+0.257~_{-1.006}^{+1.518}$ & $0.805~_{0.690}^{0.980}$ & $0.785~_{0.500}^{0.840}$ \nl
      N1023 &  149 &  $1.037~_{1.010}^{1.063}$ & $0.198~_{0.172}^{0.229}$ & $+1.660~_{+0.271}^{+2.953}$ & $1.007~_{0.878}^{1.401}$ & $0.635~_{0.500}^{0.696}$ \nl
{\it N1316} &  186 &  $0.966~_{0.951}^{0.981}$ & $0.130~_{0.114}^{0.149}$ & $-1.001~_{-2.382}^{+0.499}$ & $1.196~_{0.971}^{1.479}$ & $0.500~_{0.500}^{0.590}$ \nl
      N1399 &  410 &  $1.051~_{1.040}^{1.063}$ & $0.124~_{0.118}^{0.130}$ & $-4.387~_{-6.649}^{-2.187}$ & $0.886~_{0.784}^{0.998}$ & $0.968~_{0.908}^{0.969}$ \nl
      N1404 &  138 &  $1.017~_{0.999}^{1.037}$ & $0.137~_{0.125}^{0.152}$ & $+0.125~_{-1.470}^{+1.741}$ & $0.770~_{0.693}^{0.878}$ & $0.995~_{0.886}^{0.995}$ \nl
      N1426 &   74 &  $1.084~_{1.050}^{1.118}$ & $0.176~_{0.153}^{0.198}$ & $+0.486~_{-0.673}^{+1.660}$ & $0.965~_{0.817}^{1.205}$ & $0.995~_{0.705}^{1.000}$ \nl
      N1700 &   27 &  $1.122~_{1.059}^{1.187}$ & $0.213~_{0.163}^{0.292}$ & $+1.328~_{+0.034}^{+2.416}$ & $1.093~_{0.778}^{1.937}$ & $0.500~_{0.500}^{0.743}$ \nl
      N2300 &   82 &  $1.071~_{1.031}^{1.110}$ & $0.193~_{0.168}^{0.222}$ & $+0.881~_{-0.377}^{+2.135}$ & $0.911~_{0.772}^{1.110}$ & $0.500~_{0.500}^{0.675}$ \nl
      N2434 &  143 &  $0.945~_{0.922}^{0.968}$ & $0.189~_{0.156}^{0.224}$ & $+0.465~_{-0.723}^{+1.706}$ & $1.100~_{0.875}^{1.473}$ & $0.598~_{0.500}^{0.798}$ \nl
      N2778 &   18 &  $1.072~_{1.007}^{1.143}$ & $0.213~_{0.138}^{0.306}$ & $-0.827~_{-2.033}^{+0.597}$ & $1.291~_{0.772}^{2.567}$ & $0.595~_{0.500}^{0.800}$ \nl
      N3115 &   94 &  $1.030~_{1.002}^{1.058}$ & $0.167~_{0.146}^{0.190}$ & $+1.363~_{+0.119}^{+2.652}$ & $0.933~_{0.773}^{1.119}$ & $0.709~_{0.500}^{0.754}$ \nl
      N3377 &   86 &  $1.072~_{1.047}^{1.095}$ & $0.141~_{0.118}^{0.169}$ & $-0.680~_{-1.905}^{+0.523}$ & $1.017~_{0.826}^{1.250}$ & $0.500~_{0.500}^{0.779}$ \nl
      N3379 &   52 &  $1.095~_{1.054}^{1.129}$ & $0.172~_{0.142}^{0.208}$ & $+0.896~_{-0.307}^{+2.078}$ & $0.950~_{0.728}^{1.317}$ & $0.500~_{0.500}^{0.770}$ \nl
      N3384 &   39 &  $1.023~_{0.969}^{1.084}$ & $0.204~_{0.157}^{0.261}$ & $+1.248~_{+0.058}^{+2.285}$ & $1.067~_{0.746}^{1.640}$ & $0.500~_{0.500}^{0.737}$ \nl
      N3585 &   70 &  $1.110~_{1.071}^{1.149}$ & $0.202~_{0.173}^{0.239}$ & $+0.331~_{-0.971}^{+1.564}$ & $0.993~_{0.805}^{1.276}$ & $0.500~_{0.500}^{0.768}$ \nl
      N3610 &   29 &  $1.105~_{1.044}^{1.182}$ & $0.191~_{0.149}^{0.245}$ & $+0.154~_{-1.102}^{+1.467}$ & $0.956~_{0.747}^{1.513}$ & $0.996~_{0.557}^{1.000}$ \nl
      N3640 &   48 &  $1.080~_{1.045}^{1.116}$ & $0.158~_{0.132}^{0.186}$ & $+0.982~_{-0.250}^{+2.125}$ & $0.852~_{0.697}^{1.283}$ & $0.971~_{0.593}^{0.994}$ \nl
      N4125 &   68 &  $1.001~_{0.954}^{1.042}$ & $0.218~_{0.185}^{0.251}$ & $+0.241~_{-1.019}^{+1.560}$ & $1.007~_{0.784}^{1.378}$ & $0.500~_{0.500}^{0.814}$ \nl
{\it N4192} &   49 &  $0.884~_{0.804}^{0.970}$ & $0.366~_{0.306}^{0.425}$ & $+0.179~_{-0.910}^{+1.268}$ & $0.888~_{0.680}^{1.254}$ & $0.500~_{0.500}^{0.668}$ \nl
      N4291 &   71 &  $0.983~_{0.952}^{1.016}$ & $0.161~_{0.138}^{0.187}$ & $+0.637~_{-0.561}^{+1.770}$ & $1.072~_{0.865}^{1.348}$ & $0.500~_{0.500}^{0.941}$ \nl
{\it N4343} &   25 &  $1.026~_{0.962}^{1.096}$ & $0.246~_{0.137}^{0.325}$ & $+0.586~_{-0.622}^{+1.667}$ & $1.442~_{0.863}^{2.543}$ & $0.500~_{0.500}^{0.827}$ \nl
      N4365 &  409 &  $1.094~_{1.079}^{1.111}$ & $0.188~_{0.175}^{0.203}$ & $+0.253~_{-1.335}^{+1.851}$ & $0.941~_{0.859}^{1.047}$ & $0.757~_{0.514}^{0.795}$ \nl
      N4374 &  123 &  $0.999~_{0.977}^{1.021}$ & $0.151~_{0.133}^{0.171}$ & $+2.025~_{+0.560}^{+3.285}$ & $0.909~_{0.773}^{1.111}$ & $0.520~_{0.500}^{0.690}$ \nl
      N4406 &  131 &  $1.041~_{1.021}^{1.059}$ & $0.132~_{0.119}^{0.146}$ & $+0.283~_{-1.347}^{+1.730}$ & $0.777~_{0.690}^{0.923}$ & $0.500~_{0.500}^{0.571}$ \nl
{\it N4450} &   29 &  $0.941~_{0.850}^{1.017}$ & $0.284~_{0.189}^{0.373}$ & $-0.278~_{-1.476}^{+0.889}$ & $1.360~_{0.867}^{2.018}$ & $0.500~_{0.500}^{0.768}$ \nl
      N4458 &   33 &  $1.071~_{0.982}^{1.139}$ & $0.221~_{0.127}^{0.270}$ & $+0.887~_{-0.380}^{+1.970}$ & $0.928~_{0.556}^{1.937}$ & $0.698~_{0.500}^{0.764}$ \nl
      N4472 &  486 &  $1.095~_{1.078}^{1.113}$ & $0.250~_{0.232}^{0.269}$ & $-0.236~_{-1.865}^{+1.308}$ & $0.985~_{0.905}^{1.087}$ & $0.997~_{0.945}^{0.997}$ \nl
      N4473 &  112 &  $1.026~_{1.003}^{1.050}$ & $0.149~_{0.134}^{0.168}$ & $-0.618~_{-1.992}^{+0.983}$ & $0.896~_{0.765}^{1.052}$ & $0.500~_{0.500}^{0.667}$ \nl
      N4478 &   50 &  $1.060~_{1.022}^{1.100}$ & $0.166~_{0.136}^{0.206}$ & $+0.267~_{-1.054}^{+1.492}$ & $1.017~_{0.797}^{1.402}$ & $0.614~_{0.500}^{0.744}$ \nl
      N4486 &  655 &  $1.108~_{1.097}^{1.118}$ & $0.158~_{0.151}^{0.164}$ & $+0.290~_{-2.404}^{+2.615}$ & $0.821~_{0.764}^{0.881}$ & $0.978~_{0.951}^{0.978}$ \nl
     N4486b &   87 &  $0.978~_{0.953}^{1.001}$ & $0.142~_{0.117}^{0.175}$ & $+0.501~_{-0.746}^{+1.678}$ & $1.038~_{0.844}^{1.491}$ & $0.500~_{0.500}^{0.767}$ \nl
      N4526 &   75 &  $1.030~_{0.988}^{1.074}$ & $0.214~_{0.186}^{0.248}$ & $-0.243~_{-1.480}^{+0.993}$ & $0.815~_{0.644}^{1.037}$ & $0.999~_{0.500}^{0.999}$ \nl
{\it N4536} &   87 &  $0.766~_{0.691}^{0.847}$ & $0.388~_{0.332}^{0.439}$ & $+2.334~_{+0.951}^{+3.478}$ & $1.042~_{0.836}^{1.329}$ & $0.500~_{0.500}^{0.840}$ \nl
      N4550 &   37 &  $1.020~_{0.944}^{1.098}$ & $0.305~_{0.254}^{0.363}$ & $+0.022~_{-1.115}^{+1.218}$ & $0.909~_{0.700}^{1.238}$ & $0.500~_{0.500}^{0.688}$ \nl
      N4552 &  171 &  $1.062~_{1.043}^{1.083}$ & $0.156~_{0.142}^{0.171}$ & $+1.121~_{-0.354}^{+2.603}$ & $0.959~_{0.851}^{1.094}$ & $0.500~_{0.500}^{0.707}$ \nl
{\it N4569} &   67 &  $0.816~_{0.748}^{0.885}$ & $0.347~_{0.309}^{0.390}$ & $+0.492~_{-0.798}^{+1.667}$ & $0.988~_{0.807}^{1.181}$ & $0.973~_{0.554}^{0.988}$ \nl
{\it N4594} &  122 &  $1.065~_{1.042}^{1.089}$ & $0.152~_{0.130}^{0.176}$ & $-1.988~_{-3.310}^{-0.430}$ & $0.848~_{0.687}^{1.121}$ & $0.995~_{0.627}^{1.000}$ \nl
      N4621 &  126 &  $1.079~_{1.059}^{1.099}$ & $0.128~_{0.116}^{0.141}$ & $-1.611~_{-3.073}^{-0.130}$ & $0.916~_{0.773}^{1.119}$ & $0.539~_{0.500}^{0.654}$ \nl
      N4649 &  378 &  $1.096~_{1.081}^{1.111}$ & $0.185~_{0.171}^{0.200}$ & $-1.401~_{-2.909}^{+0.251}$ & $0.867~_{0.794}^{0.946}$ & $0.961~_{0.810}^{0.963}$ \nl
      N4660 &   74 &  $0.962~_{0.935}^{0.992}$ & $0.148~_{0.117}^{0.183}$ & $+0.899~_{-0.330}^{+2.148}$ & $1.091~_{0.832}^{1.445}$ & $0.500~_{0.500}^{0.767}$ \nl
      N4881 &   84 &  $0.915~_{0.868}^{0.959}$ & $0.276~_{0.225}^{0.337}$ & $+0.505~_{-0.497}^{+1.683}$ & $1.321~_{1.026}^{1.770}$ & $0.500~_{0.500}^{0.999}$ \nl
      N5018 &   10 &  $0.931~_{0.848}^{1.013}$ & $0.187~_{0.111}^{0.309}$ & $+0.193~_{-1.145}^{+1.340}$ & $1.714~_{0.965}^{3.228}$ & $0.500~_{0.500}^{0.773}$ \nl
      N5061 &   84 &  $1.099~_{1.068}^{1.127}$ & $0.190~_{0.143}^{0.242}$ & $-0.473~_{-1.602}^{+0.723}$ & $1.231~_{0.907}^{1.812}$ & $0.500~_{0.500}^{0.800}$ \nl
      N5845 &   28 &  $1.127~_{1.044}^{1.215}$ & $0.285~_{0.211}^{0.364}$ & $+0.543~_{-0.617}^{+1.569}$ & $1.272~_{0.878}^{1.851}$ & $0.500~_{0.500}^{0.733}$ \nl
      N5846 &  781 &  $1.080~_{1.066}^{1.094}$ & $0.249~_{0.232}^{0.266}$ & $-2.895~_{-4.551}^{-1.318}$ & $1.167~_{1.070}^{1.261}$ & $0.895~_{0.692}^{0.935}$ \nl
      N7192 &   95 &  $1.068~_{1.029}^{1.112}$ & $0.262~_{0.215}^{0.313}$ & $-0.196~_{-1.338}^{+0.990}$ & $1.115~_{0.882}^{1.446}$ & $0.500~_{0.500}^{0.810}$ \nl
      N7457 &   48 &  $1.024~_{0.991}^{1.056}$ & $0.147~_{0.113}^{0.206}$ & $+0.673~_{-0.461}^{+1.802}$ & $1.288~_{0.857}^{2.449}$ & $0.500~_{0.500}^{0.770}$ \nl
     IC4889 &   48 &  $1.024~_{0.992}^{1.056}$ & $0.147~_{0.113}^{0.203}$ & $+0.673~_{-0.530}^{+1.800}$ & $1.288~_{0.875}^{2.521}$ & $0.500~_{0.500}^{0.800}$ \nl
\enddata
\end{deluxetable}

\begin{deluxetable}{llcrlcrcc}
\footnotesize
\tablecaption{Moments of Combined Samples}
\tablewidth{0pt}
\tablehead{
\colhead{Sub-Sample}  &
\colhead{Range}       &
\colhead{N$_{\rm G}$} &
\colhead{N$_{\rm GC}$}& 
\colhead{$V-I$ peak}  &
\colhead{Scale}       &
\colhead{Skewness}    &
\colhead{Tail}        &
\colhead{DIP}         } 
\startdata
Faint  & M$_V>-20.70$ & 15 &  803 & $1.019~_{1.008}^{1.029}$ & $0.182~_{0.173}^{0.192}$ & $+1.663~_{+0.302}^{+3.005}$ & $1.001~_{0.941}^{1.065}$ & $0.742~_{0.500}^{0.815}$ \nl
Bright & M$_V<-21.75$ & 13 & 2737 & $1.077~_{1.072}^{1.083}$ & $0.172~_{0.168}^{0.176}$ & $-1.397~_{-3.383}^{+0.518}$ & $0.884~_{0.857}^{0.913}$ & $0.954~_{0.781}^{0.963}$ \nl
&&&&&&&&\nl
Low $\sigma$ & log$(\sigma) < 2.3$ & 16 &  872 & $1.041~_{1.031}^{1.051}$ & $0.188~_{0.178}^{0.197}$ & $+1.243~_{-0.118}^{+2.509}$ & $1.022~_{0.964}^{1.080}$ & $0.759~_{0.500}^{0.904}$ \nl
High $\sigma$ & log$(\sigma) >2.4$ & 12 & 2992 & $1.075~_{1.070}^{1.080}$ & $0.173~_{0.169}^{0.177}$ & $-0.455~_{-2.586}^{+1.662}$ & $0.879~_{0.855}^{0.907}$ & $0.996~_{0.959}^{0.996}$ \nl
&&&&&&&&\nl
Low Mg2 & Mg2 $< 0.30$ & 18 & 1496 & $1.069~_{1.061}^{1.078}$ & $0.205~_{0.197}^{0.214}$ & $-0.104~_{-1.894}^{+1.786}$ & $1.039~_{0.987}^{1.094}$ & $0.791~_{0.562}^{0.851}$ \nl
High Mg2 & Mg2 $> 0.34$ & 15 & 2597 & $1.070~_{1.064}^{1.075}$ & $0.170~_{0.166}^{0.175}$ & $-0.292~_{-2.258}^{+1.618}$ & $0.888~_{0.859}^{0.916}$ & $0.985~_{0.936}^{0.985}$ \nl
&&&&&&&&\nl
Low $\rho$ & log$(\rho) < 1.0$ & 24 & 2202 & $1.057~_{1.050}^{1.064}$ & $0.200~_{0.194}^{0.207}$ & $+1.306~_{-0.364}^{+2.994}$ & $1.019~_{0.977}^{1.057}$ & $0.967~_{0.780}^{0.974}$ \nl
High $\rho$ & log$(\rho) > 1.0$ & 17 & 3441 & $1.069~_{1.064}^{1.074}$ & $0.170~_{0.166}^{0.173}$ & $-0.262~_{-2.304}^{+1.903}$ & $0.887~_{0.862}^{0.909}$ & $0.999~_{0.948}^{0.999}$ \nl
\enddata
\end{deluxetable}

\clearpage

\onecolumn

\begin{figure}
\plotfiddle{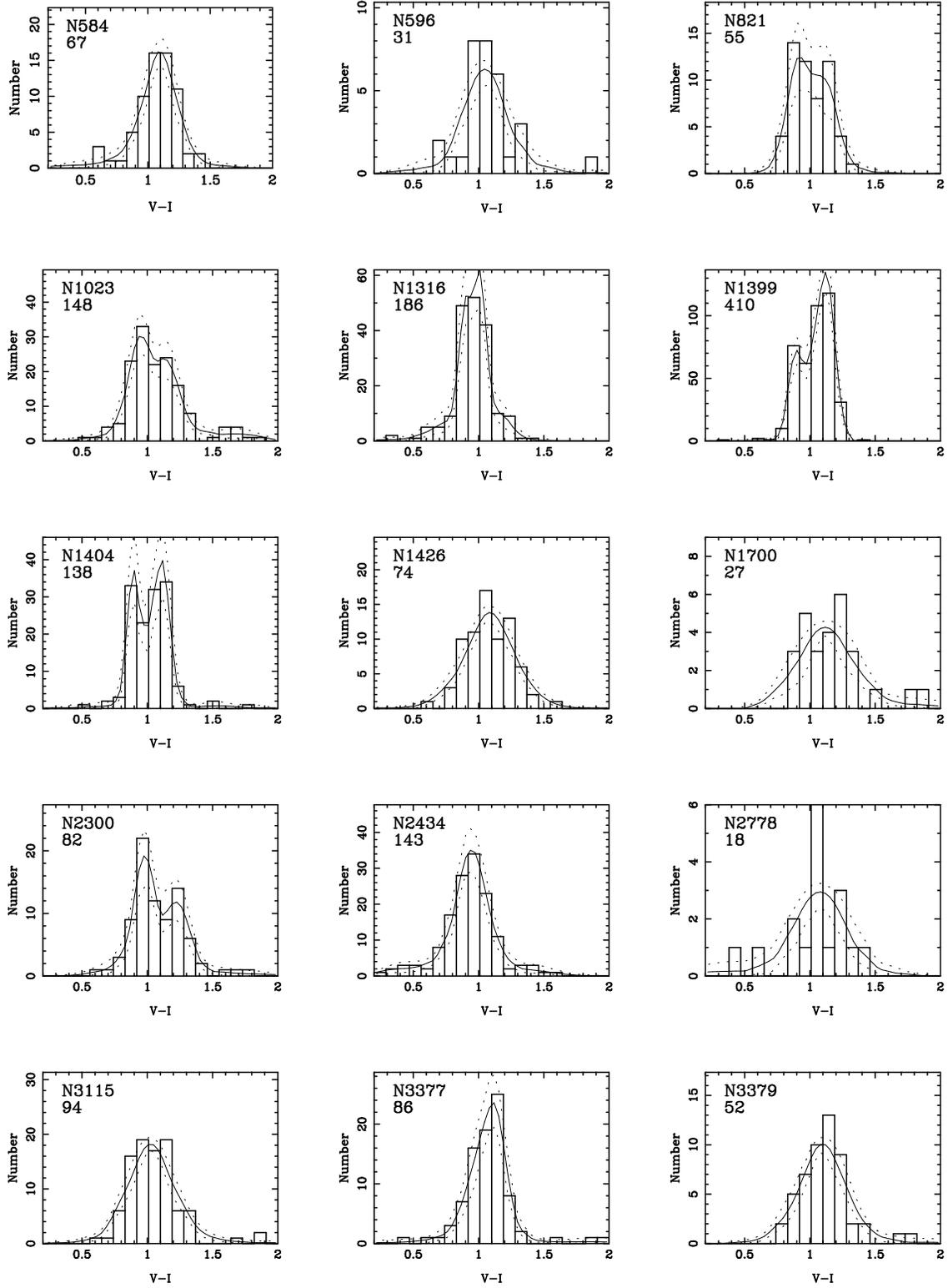}{550pt}{0}{80}{80}{-240pt}{-20pt}
\caption{Histogram of the globular cluster system's $V-I$ colors for
all of the galaxies. The solid line represents a density estimate and
the dotted lines are the 90\% confidence bands. The total number of
globular clusters appears under the galaxy name.}
\end{figure}

\begin{figure}
\figurenum{1}
\plotfiddle{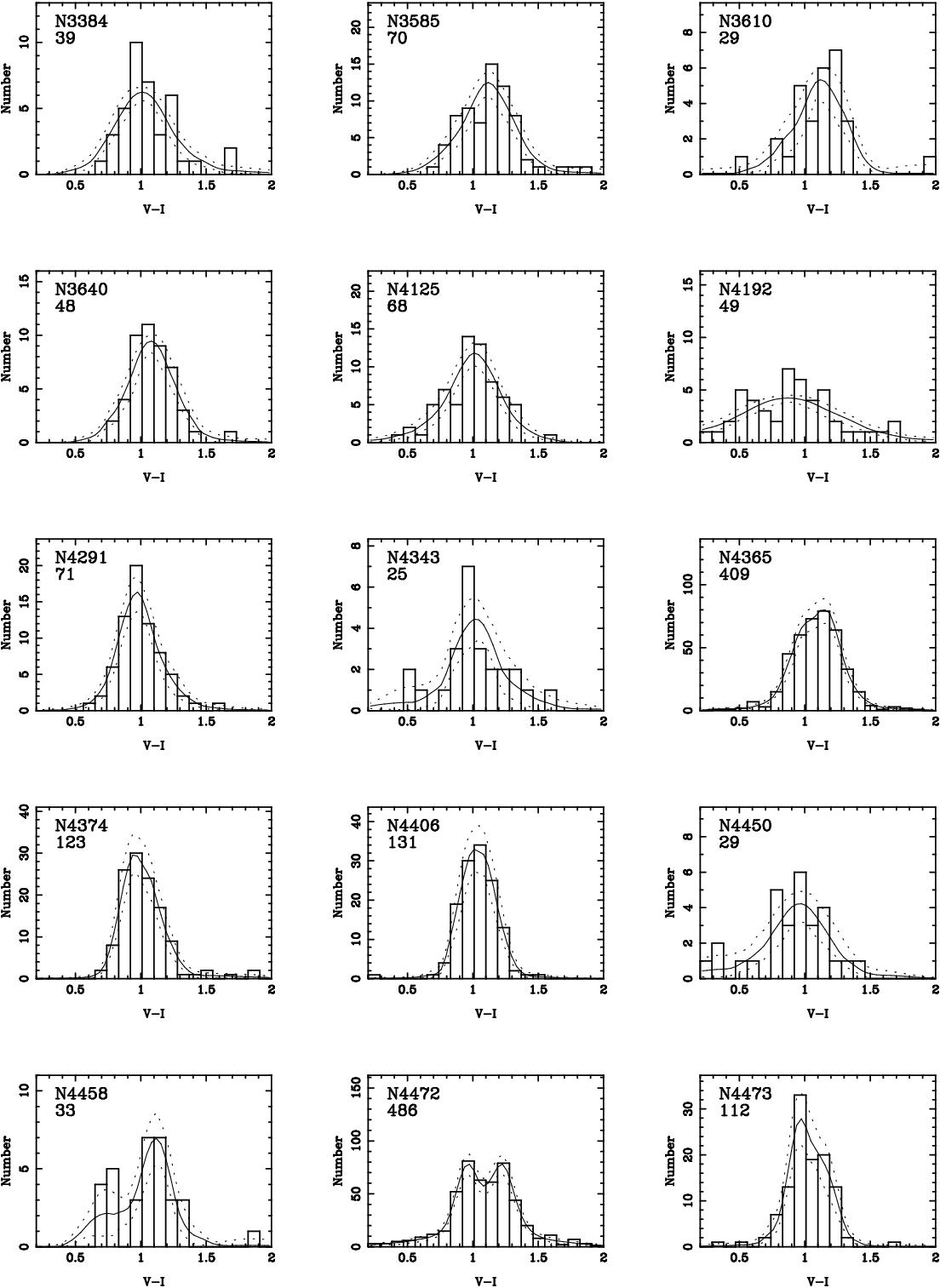}{550pt}{0}{80}{80}{-240pt}{-20pt}
\caption{continued.}
\end{figure}

\begin{figure}
\figurenum{1}
\plotfiddle{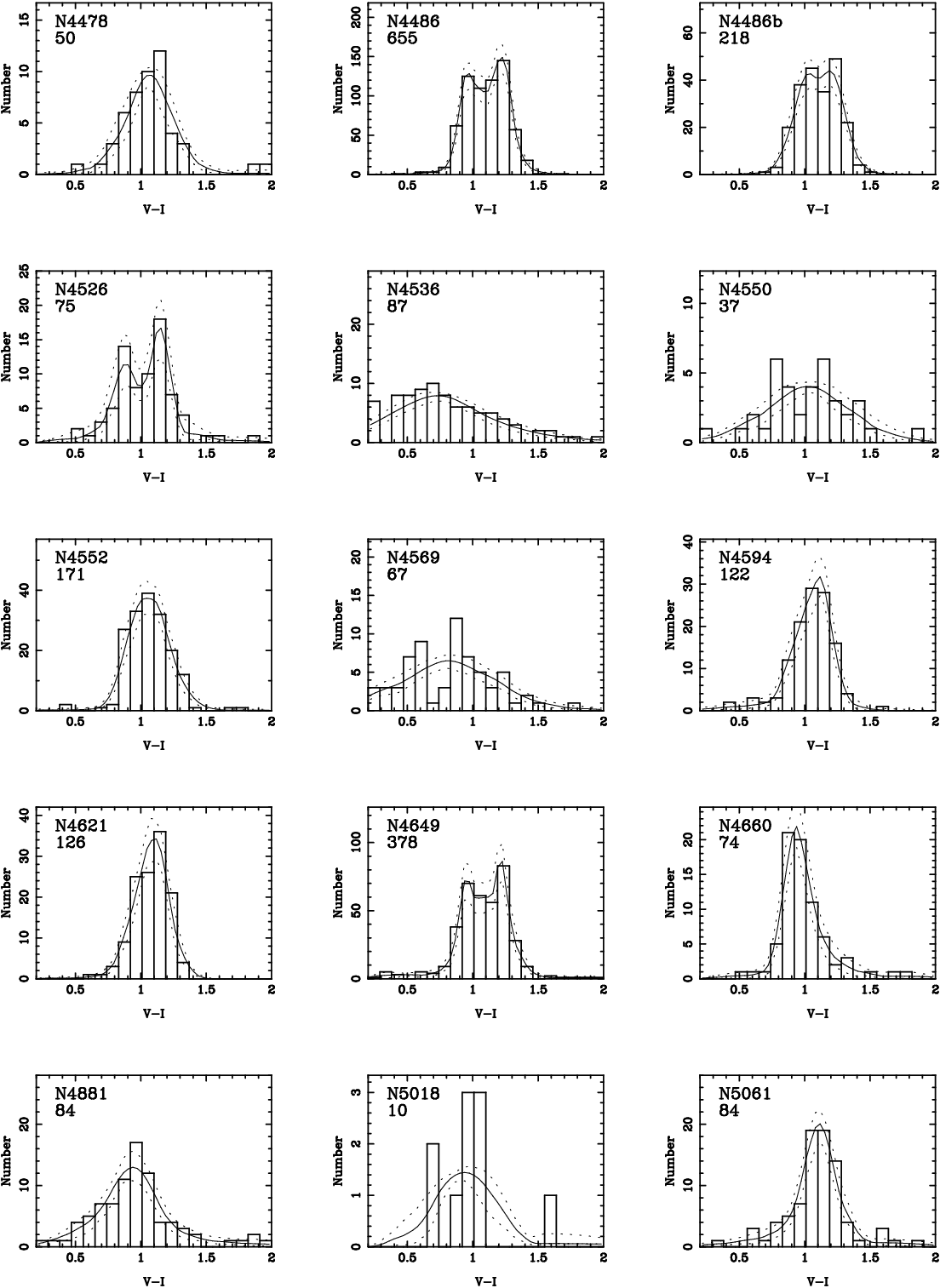}{550pt}{0}{80}{80}{-240pt}{-20pt}
\caption{continued.}
\end{figure}

\begin{figure}
\figurenum{1}
\plotfiddle{gebhardt.fig1d.ps}{550pt}{0}{80}{80}{-240pt}{-20pt}
\caption{continued.}
\end{figure}

\begin{figure}
\plotfiddle{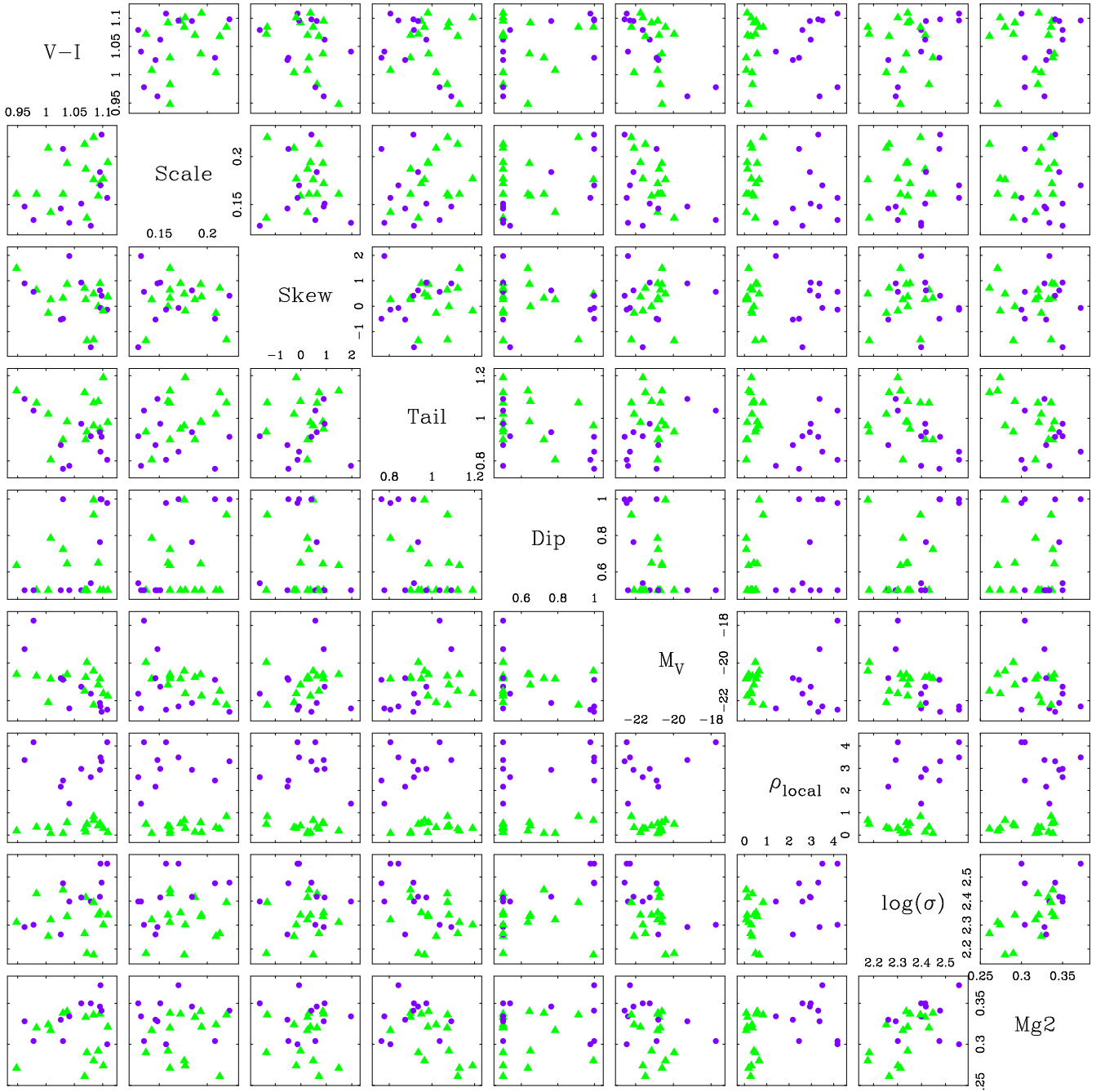}{520pt}{0}{90}{90}{-270pt}{20pt}
\caption{Plots of various globular cluster systems properties and
parent galaxy properties. The triangles represent field galaxies and
the circles represent cluster galaxies. The names along the diagonal
are the variable which is plotted along that particular row and
column. Each plot is repeated twice here as the plots in the
lower-left triangle and the upper-right triangle represent the same
points with the axes inverted.}
\end{figure}

\begin{figure}
\plotfiddle{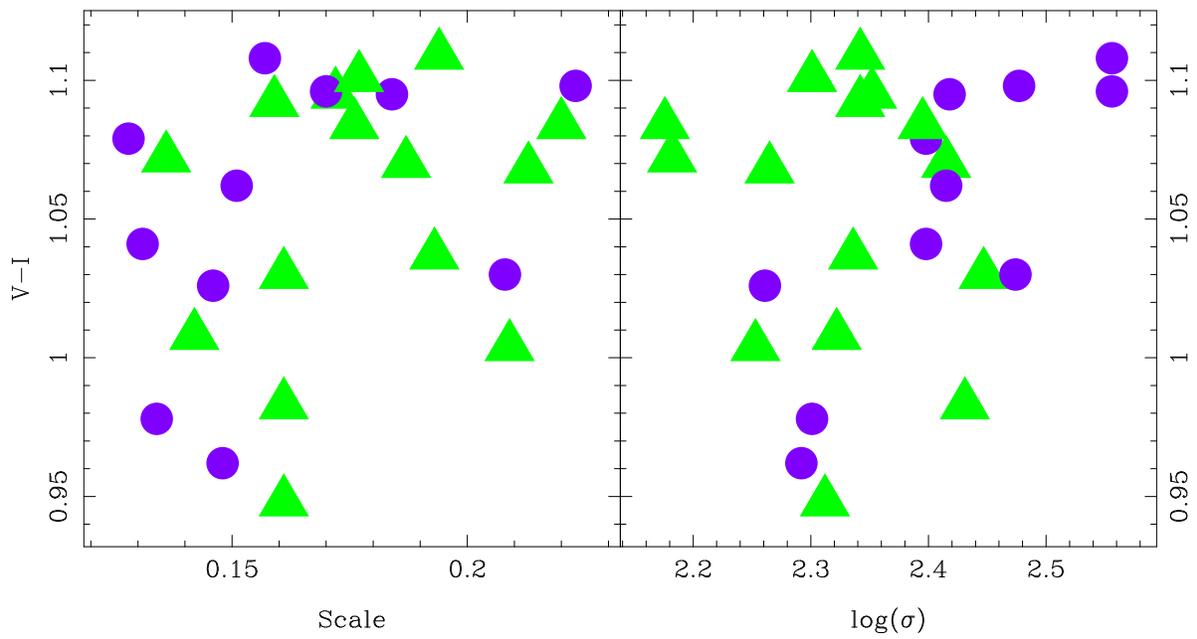}{520pt}{0}{90}{90}{-270pt}{100pt}
\caption{Scale and log($\sigma$) versus the $V-I$ color of the
globular cluster systems. The colors of the points represent the same
as in the previous plot.}
\end{figure}

\begin{figure}
\plotfiddle{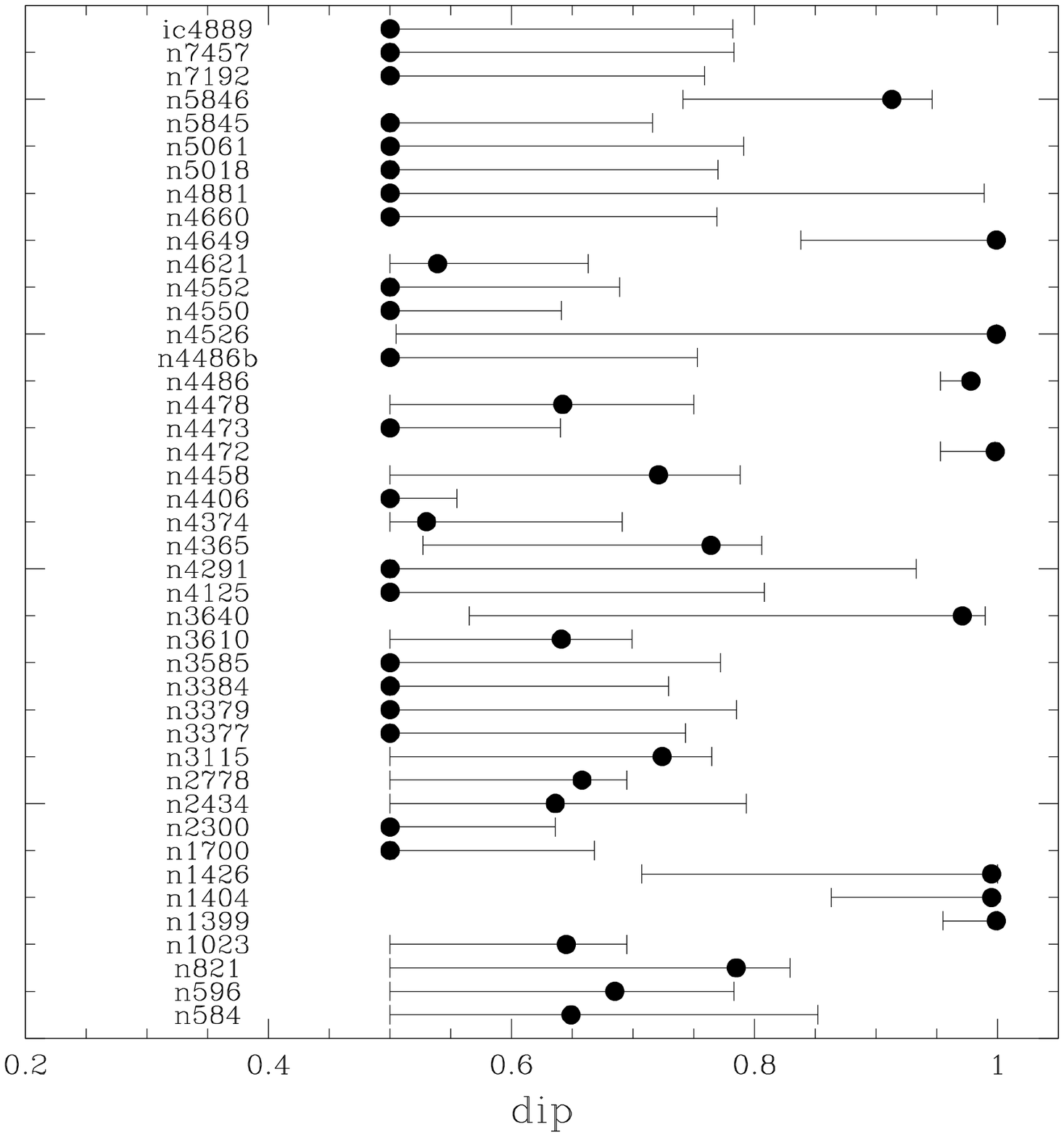}{520pt}{0}{90}{90}{-270pt}{-100pt}
\caption{The distribution of the DIP probability with it's 68\%
confidence band for all of the galaxies. The DIP probability is
bounded by 0.5 and 1.0.}
\end{figure}

\begin{figure}
\plotfiddle{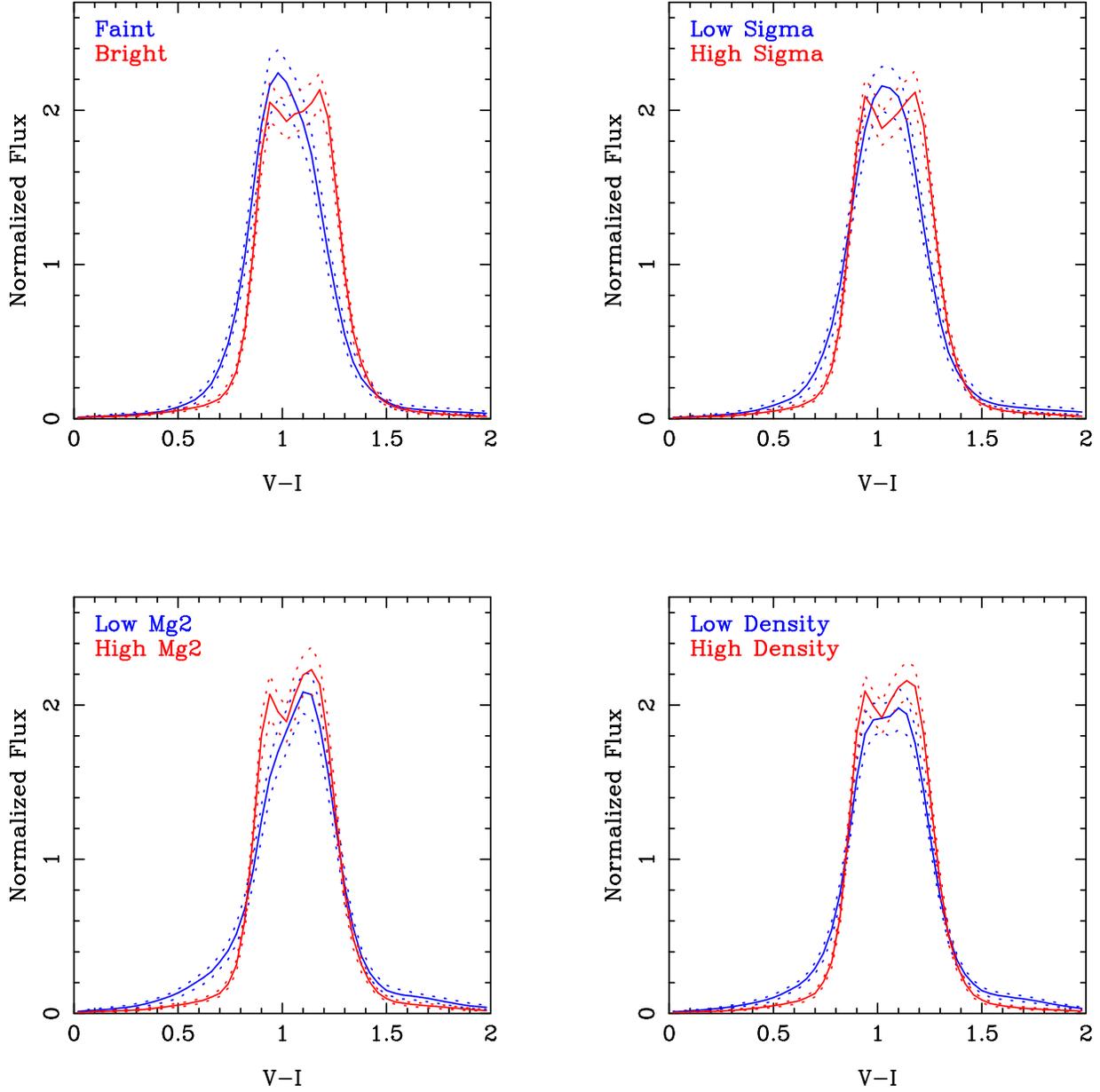}{520pt}{0}{90}{90}{-270pt}{20pt}
\caption{$V-I$ distributions and their 90\% confidence bands for the
globular clusters combined across galaxies; we use two bins for each
of the named galaxy properties. Each curve represents an adaptive
kernel estimate of the density with unity area.}
\end{figure}

\end{document}